\documentclass[letterpaper,twocolumn,prx,aps,superscriptaddress,floatfix]{revtex4}
\usepackage{mathptmx}

\usepackage[latin9]{inputenc}
\usepackage{amsmath}
\usepackage{amssymb}
\usepackage{graphicx}
\usepackage{esint}

\makeatletter

\pdfpageheight\paperheight
\pdfpagewidth\paperwidth

\providecommand{\tabularnewline}{\\}

\@ifundefined{textcolor}{}
{%
 \definecolor{BLACK}{gray}{0}
 \definecolor{WHITE}{gray}{1}
 \definecolor{RED}{rgb}{1,0,0}
 \definecolor{GREEN}{rgb}{0,1,0}
 \definecolor{BLUE}{rgb}{0,0,1}
 \definecolor{CYAN}{cmyk}{1,0,0,0}
 \definecolor{MAGENTA}{cmyk}{0,1,0,0}
 \definecolor{YELLOW}{cmyk}{0,0,1,0}
}

\usepackage{xcolor}\usepackage{soul}

\newcommand{\ket}[1]{\ensuremath{\left|#1\right\rangle}}
\definecolor{blue}{rgb}{0,0,1}
\definecolor{red}{rgb}{1,0,0}
\definecolor{green}{rgb}{0,1,0}

\makeatother

\begin{document}

\title{Converting quasiclassical states into arbitrary Fock state superpositions in a superconducting cavity}

\author{W.~Wang}
\author{L.~Hu}
\author{Y.~Xu}
\author{K.~Liu}
\author{Y.~Ma}
\affiliation{Center for Quantum Information, Institute for Interdisciplinary Information
Sciences, Tsinghua University, Beijing 100084, China}
\author{Shi-Biao Zheng}
\email{t96034@fzu.edu.cn}
\affiliation{Department of Physics, Fuzhou University, Fuzhou 350116, China}
\author{R.~Vijay}
\affiliation{Department of Condensed Matter Physics and Materials Science, Tata
Institute of Fundamental Research, Homi Bhabha Road, Mumbai 400005,
India}
\author{Y.~P.~Song}
\affiliation{Center for Quantum Information, Institute for Interdisciplinary Information
Sciences, Tsinghua University, Beijing 100084, China}
\author{L.-M. Duan}
\email{lmduan@umich.edu}
\affiliation{Center for Quantum Information, Institute for Interdisciplinary Information
Sciences, Tsinghua University, Beijing 100084, China}
\affiliation{Department of Physics, University of Michigan, Ann Arbor, Michigan
48109, USA}
\author{L.~Sun}
\email{luyansun@tsinghua.edu.cn}
\affiliation{Center for Quantum Information, Institute for Interdisciplinary Information
Sciences, Tsinghua University, Beijing 100084, China}

\begin{abstract}
We propose and experimentally demonstrate a new method to generate
arbitrary Fock-state superpositions in a superconducting quantum circuit,
where a qubit is dispersively coupled to a microwave cavity mode without
the need of fine-frequency tuning. Here the qubit is used to conditionally modulate the probability amplitudes of the Fock state components of a coherent state to those of the desired superposition state, instead of pumping photons one by one into the cavity as in previous schemes. Our method does not require the adjustment of the qubit frequency during the cavity state preparation, and is more robust to noise and accumulation of experimental errors compared to previous ones. Using the method, we experimentally generate high-fidelity phase eigenstates under various Hilbert-space dimensions and squeezed states, which are useful for quantum walk and high-precision measurements.
\end{abstract}
\maketitle
\vskip 0.5cm

\narrowtext

The creation and manipulation of complex superposition states of a
harmonic oscillator (e.g., a light field) is an important subject
in quantum optics~\cite{Buzek1995}, and has various applications
in quantum technologies. In contrast to nonlinear quantum systems
(e.g., atoms), a harmonic oscillator possesses an infinite-dimensional
Hilbert space where the energy is linear in the quantum number. Due
to this feature, a harmonic oscillator has extremely rich nonclassical
effects and serves as an ideal candidate for investigation of fundamental
quantum theory. In addition to fundamental interest, it plays important
roles in many quantum computation protocols. On one hand, it acts
as a data bus to link different qubits, as in the ones with trapped
ions~\cite{Cirac1995}, cavity quantum electrodynamics (QED)~\cite{Raimond2001},
and circuit QED~\cite{Sillanpaa2007,Majer2007,DiCarlo2009}. It can
also be used as a quantum memory to store quantum bits (qubits), offering
the possibility to construct a quantum processor with fewer resources~\cite{Santos2005}
and to realize quantum error correction~\cite{LeghtasPRL2013,Ofek2016,Michael2016}.
On the other hand, harmonic oscillators are the basic building block
for implementation of quantum computation based on continuous variables~\cite{Lloyd1999,Gottesman2001}.
In addition, the nonclassical effects of harmonic oscillators have
applications in other quantum technologies, e.g., when a quantum state
exhibits squeezing, i.e., the reduction of the fluctuation in one
certain quadrature below the vacuum level, it can be used in high-precision
measurement~\cite{Caves1980}. Many of these applications rely on
the ability to generate an arbitrary superposition state of a harmonic
oscillator, however, its linear energy spectrum prevents one from
doing so by using classical drives only; the application of a classical
drive produces a displacement in phase space without enhancing the
nonclassical behaviors.

Previous methods employed a step-by-step algorithm~\cite{Vogel1993,Law1996},
where in each step the quantum state of a two-level system is manipulated
in a controllable manner and then transferred to the harmonic oscillator,
increasing the quantum number of the largest Fock state component in the superposition by $1$. After $N$ steps, the oscillator evolves from the ground state to any superposition
of the first $N+1$ number states. Such a technique was first demonstrated
on the harmonic motion of a trapped ion trap~\cite{Ben-Kish2003},
and then implemented in a superconducting resonator~\cite{Hofheinz2008,Hofheinz2009}
using a qubit with fast frequency tunability to coherently pump photons
from external microwave drive pulses into the resonator. More recently,
it was shown that $2N$ selective number-dependent arbitrary phase
gates, together with $2N+1$ displacement operations, can be used
to construct such superposition states~\cite{Krastanov2015}. Following
this proposal, a one-photon Fock state was created in a circuit QED
system~\cite{Heeres2015}. With a similar setup, the Gradient Ascent Pulse Engineering method~\cite{Khaneja2005,deFouquieres2011} was recently used for creating the six-photon state and four-component cat states~\cite{Heeres2016}. In the context of cavity QED, Fock states and cat states of a cavity were conditionally generated with Rydberg atoms dispersively coupled to the cavity mode~\cite{Deleglise}.

Here we propose and experimentally demonstrate a distinct
method to synthesize any superposition of Fock states for the field
stored in a cavity that is dispersively coupled to a qubit. Our method
relies on modulating the probability amplitudes associated with the
superposed Fock states in an initial coherent state, a superposition
of photon number states with a Poissonian distribution, to those of the desired
target state. Coherent states are quasiclassical states characterized by a complex amplitude~\cite{Deleglise}, and can be generated by using a classical drive. The photon-number-dependent shift in the qubit frequency, arising from the dispersive qubit-cavity coupling, enables one to use carefully tailored microwave drive signals to individually control
the qubit transition amplitude associated with each Fock state. The
subsequent measurement of the qubit transition conditionally projects
the cavity to a superposition of Fock states, with the probability
amplitude of each state component being proportional to the associated
qubit transition amplitude, controllable by the amplitude and phase
of the corresponding drive signal. Unlike the previous methods, it
is unnecessary to individually pump photons into the cavity one by
one, and the modulating drive signals can be simultaneously applied
and incorporated into a single pulse. Furthermore, during the preparation process the qubit-cavity detuning needs not to be adjusted, which is important for quantum state engineering in circuit QED systems with three-dimensional cavities~\cite{Heeres2015,Paik,Vlastakis,SunNature},
where it is difficult to adjust the qubit frequency. We implement
the experiment in a superconducting circuit, where a three-dimensional
cavity is coupled to a transmon qubit. We analyze the produced states
by measuring the Wigner functions, showing good agreement with the
desired ones.

The system under investigation consists of a qubit dispersively coupled
to a cavity and driven by a classical microwave containing $N+1$
frequency components. In the interaction picture, the Hamiltonian
for the total system is
\begin{equation}
H_{I}=-\chi_{qc}a^{\dagger}a\left|e\right\rangle \left\langle e\right|+\left[\sum_{n=0}^{N}\Omega_{n}e^{i(\delta_{n}t+\phi_{n})}\left|e\right\rangle \left\langle g\right|+h.c.\right],
\label{eq:Hamiltonian}
\end{equation}
where $\left|e\right\rangle $ and $\left|g\right\rangle $ denote
the excited and ground states of the qubit, $a^{\dagger}$ and $a$
are the creation and annihilation operators of the photonic field
in the cavity, $\chi_{qc}$ is the qubit's frequency shift induced by per photon in the cavity due to the dispersive qubit-cavity coupling, $\delta_{n}$ is the detuning between the qubit and the
$n$th driving component with a Rabi frequency $\Omega_{n}$ and a
phase $\phi_{n}$, and $h.c.$ denotes the Hermitian conjugate. The
qubit is initially in the ground state $\left|g\right\rangle $, and
the cavity in a coherent state
\begin{equation}
\left|\alpha\right\rangle =\sum_{n=0}^{\infty}c_{n}\left|n\right\rangle,
\end{equation}
where $c_{n}=\exp(-\left|\alpha\right|^{2}/2)\alpha^{n}/\sqrt{n!}$
is the probability amplitude for having $n$ photons. The photon-number-dependent
shift of the qubit transition frequency enables individually modulating
the probability amplitudes of the superposed Fock states in $\left|\alpha\right\rangle $
with a carefully tailored microwave drive addressing the qubit. Our
aim is to convert such a quasiclassical state to a target quantum
state 
\begin{equation}
\left|\psi_{d}\right\rangle =\sum_{n=0}^{N}d_{n}\left|n\right\rangle,
\end{equation}
with the desired complex amplitudes $d_{n}$ for the corresponding
Fock state components.

\begin{figure}
\includegraphics{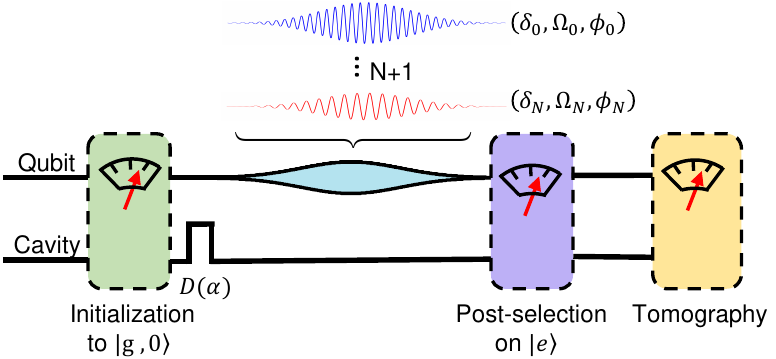} \caption{(Color online) \textbf{Experimental sequence.} The experiment starts with the application of a microwave pulse to the cavity, corresponding
to a phase-space displacement operation $D(\alpha)$, which transforms
the cavity from the ground state $|0\rangle$ to the coherent state
$|\alpha\rangle$. Then the qubit, initially in the ground state $|g\rangle$,
is driven by a classical microwave pulse comprising $N+1$ frequency
components, with the $n$th component resonant with the qubit transition
conditional on the cavity's photonic field being in the Fock state
with $n$ photons. This corresponds to simultaneous implementation
of $N+1$ conditional qubit rotations on the Bloch sphere:$\prod_{n=0}^{N}R_{e_{n}|n}(\beta_{n})$.
Subsequent measurement of the qubit in the excited state $|e\rangle$
collapses the cavity to a superposition of $N+1$ Fock states, whose
amplitudes are controllable by the parameters of the corresponding
microwave components. A final Wigner tomography is used to examine the produced states.}
\label{fig:set} 
\end{figure}

Under the condition $\delta_{n}=n\chi_{qc}$ and $\Omega_{n}\ll\chi_{qc}$,
the $n$th component resonantly drives the qubit transition when the
cavity is in the Fock state $\left|n\right\rangle $, but does not
affect the qubit state if the cavity is in any other Fock state due
to large detunings. The dynamics governed by $H_{I}$ enables individually
controlling the transition in each subspace \{$\left|g,n\right\rangle $,
$\left|e,n\right\rangle $\} (the notation denotes $\left|qubit,cavity\right\rangle $);
there is no interaction between different subspaces. After an interaction
time $\tau$ the qubit-cavity system evolves to
\begin{equation}
\left|\psi_{qc}(\tau)\right\rangle =\sum_{n=0}^{\infty}c_{n}\left[\cos\beta_{n}\left|g,n\right\rangle -ie^{i(\phi_{n}+n\chi_{qc}\tau)}\sin\beta_{n}\left|e,n\right\rangle \right],
\end{equation}
where $\beta_{n}=\int_{0}^{\tau}\Omega_{n}dt$ is the pulse area associated
with the $n$th frequency component. We take $\beta_{n}=0$ for $n>N$,
where $N$ is the maximum photon number in the desired target state.
The detection of the qubit in the state $\left|e\right\rangle $ projects
the cavity to the superposition state
\begin{equation}
\left|\psi_{c}\right\rangle =\mathcal{N}\sum_{n=0}^{N}c_{n}e^{i(\phi_{n}+n\chi_{qc}\tau)}\sin\beta_{n}\left|n\right\rangle ,
\end{equation}
where $\mathcal{N=}\left[\sum_{n=0}^{N}\left|c_{n}\right|^{2}\sin^{2}\beta_{n}\right]^{-1/2}$
is a normalization factor. The final cavity state $\left|\psi_{c}\right\rangle $
is equivalent to the desired state $\left|\psi_{d}\right\rangle $
when the parameters $\beta_{n}$ and $\phi_{n}$ are appropriately
chosen so that $d_{n}=\mathcal{N}c_{n}\sin\beta_{n}e^{i(n\chi_{qc}\tau+\phi_{n})}$.
To maximize the success probability $P=1/\mathcal{N}^{2}$, we numerically
optimize the parameters $\left|\alpha\right|$ and $\beta_{n}$, with
the optimal values depending upon the desired state.

Our experiment is implemented with a superconducting circuit, where
one transmon qubit is dispersively coupled to two three-dimensional
cavities~\cite{Wallraff,Paik,Vlastakis,SunNature,Liu2016}. The qubit, with a transition
frequency $\omega_{q}/2\pi=5.345$~GHz, has an energy relaxation time
$T_{1}=13.0$~$\mu$s and a pure dephasing time $T_{\phi}=13.8$~$\mu$s.
Cavity~1, with a frequency $\omega_{s}/2\pi=8.230$~GHz and a photon
lifetime $\tau_{s}=80$~$\mu$s, is used for storage of the photonic
state. Cavity~2, with a frequency $\omega_{r}/2\pi=7.290$~GHz and
a photon lifetime $\tau_{r}=44$~ns, is connected to a Josephson
parametric amplifier~\cite{Hatridge,Roy,Kamal,Murch}, allowing for
a high-fidelity (nearly-unity) and quantum non-demolition single-shot
readout of the qubit state. The average readout photon number in the
readout cavity is about five and the readout time is $240$~ns. For
simplicity, the term ``cavity" refers to the cavity~1.
The dispersive frequency shift is $\chi_{qc}/2\pi=-1.44$~MHz.
The experimental apparatus and readout properties can be found in the Supplementary Material and Ref.~\cite{Liu2016}.

The experimental sequence is outlined in Fig.~\ref{fig:set}. The
initialization of the system to $\left|g,0\right\rangle $ is realized
by the post-selection on the qubit's ground state and the subsequent
cavity parity measurement. The initial coherent state of the cavity
is achieved by application of a microwave pulse, which produces a
displacement operation $D(\alpha)$, transforming the cavity from
the ground state $\left|0\right\rangle $ to $\left|\alpha\right\rangle $.
This is followed by driving the qubit, initially in the ground state
$\left|g\right\rangle $, with a pulse with $N+1$ frequency components
resonant with the corresponding qubit-cavity transition frequencies,
simultaneously implementing $N+1$ conditional qubit rotations: $\prod_{n=0}^{N}R_{e_{n}|n}(\beta_{n})$, where $R_{e_n|n}(\beta _n)$ represents the qubit rotation on the Bloch sphere by an angle $\beta _n$ around the $e_n$ direction conditional on the cavity containing $n$ photons, with $e_n$ being on the equatorial plane with the azimuth angle $\phi _n+n\chi_{qc}\tau$. In order to have enough selectivity, each frequency component in the qubit pulse simply has a broad Gaussian envelop truncated to 4$\sigma=1.44~\mu$s ($\sigma_{f}=0.44$~MHz). Post-selecting on the excited qubit state, we obtain the desired cavity state. After preparation of the desired cavity state, the qubit is disentangled with the cavity and can be used to measure the Wigner function of the cavity for examination of the produced state. Following a technique devised by Lutterbach and Davidovich~\cite{Lutterbach1997} and demonstrated in cavity QED~\cite{Bertet2002} and circuit QED~\cite{Vlastakis,SunNature,Liu2016},
the Wigner quasiprobability distribution $W(\beta)$ is measured by
a combination of the cavity's displacement operation $D(-\beta)$
and qubit Ramsey-type measurement, where a conditional cavity $\pi$
phase shift $C(\pi)$ is sandwiched in between two unconditional qubit
rotations $R_{y}(\pi/2)$.

As an example, we prepared the truncated phase states~\cite{Pegg1997}
\begin{equation}
\left|\theta_{\mathrm{N,k}}\right\rangle =\frac{1}{\sqrt{N+1}}\sum_{n=0}^{N}e^{in\theta_{\mathrm{N,k}}}\left|n\right\rangle
\end{equation}
with $N=5$ to $7$, where $\theta_{\mathrm{N,k}}=2k\pi/(N+1)$. Apart from the fundamental interest, such states are useful for implementation of quantum walk~\cite{Sanders2003}. The Wigner functions of the ideal states and the produced states with $\theta_{\mathrm{N,k}}=0$ are shown in the first and second rows of Fig.~\ref{fig:fock}, respectively.
We note that, mainly due to the decoherence of the system (see Supplementary Material), the contrast of the measured Wigner function is reduced by a factor $R$: $W_{m}(\beta)=RW_{r}(\beta)$,
where $W_{m}(\beta)$ and $W_{r}(\beta)$ are the measured and real
Wigner functions of the produced state~\cite{VlastakisBell}. The
reduction factor is $R=\int W_{m}(\beta)d^{2}\beta$. To make the
results physically meaningful, we normalize the measured Wigner functions,
taking $W_{m}(\beta)/R$ as the real Wigner functions of the produced
states. The shapes of the Wigner functions of the produced states
agree very well with those of the ideal ones, implying that the quasiclassical
states have been converted into the desired quantum states with high
accuracy. The amplitudes $\alpha$ of the initial coherent states,
Wigner function reduction factors $R$, success probabilities $P$,
and fidelities $F$ between the produced states and desired ones are
detailed in Table~\ref{T:fock}. Here the fidelity is defined as
$F=\left\langle \theta_\mathrm{{N,k}}\right|\rho_{p}\left|\theta_\mathrm{{N,k}}\right\rangle $,
where $\rho_{p}$ is the density operator of the produced state reconstructed
by least-square regression using a maximum likelihood estimation~\cite{Smolin2012,VlastakisBell}. The infidelity is mainly due to the decoherence of the system and imperfection of the conditional qubit rotations. However, since the desired cavity state is obtained by post-selection on the qubit's excited state $\ket{e}$, its fidelity is actually insensitive to the decoherence of the qubit (see Supplementary Material) as demonstrated by the measured results in Table~\ref{T:fock}.


\begin{figure*}
\includegraphics{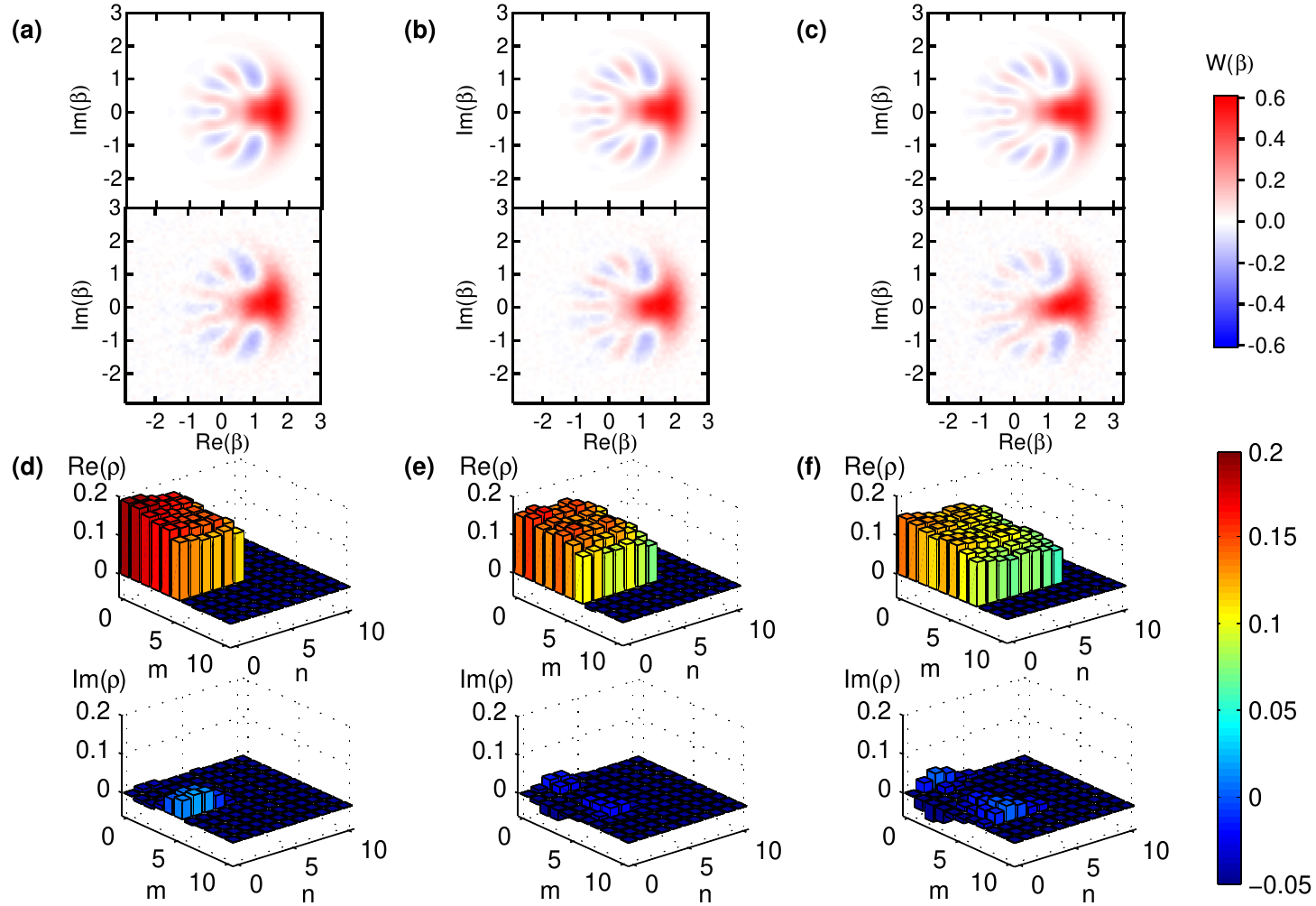} \caption{(Color online) \textbf{Wigner tomography and density matrix reconstruction of the truncated phase states with $\theta_{\mathrm{N,k}}=0$.} The Wigner functions of the ideal and produced truncated phase states, with $N=5$ to $7$, are displayed in the first and second rows, respectively. Due to the readout errors, the measured Wigner functions $W_{m}(\beta)$ are reduced by a factor $R$ compared to the real ones, where $R=\int W_{m}(\beta)d^{2}\beta$. We take $W_{m}(\beta)/R$ as the real Wigner functions of the produced states. The real and imaginary parts of the density matrices, obtained from the real Wigner functions, are shown in the third and fourth rows.}
\label{fig:fock} \vspace{-6pt}
\end{figure*}

\begin{figure*}
\includegraphics{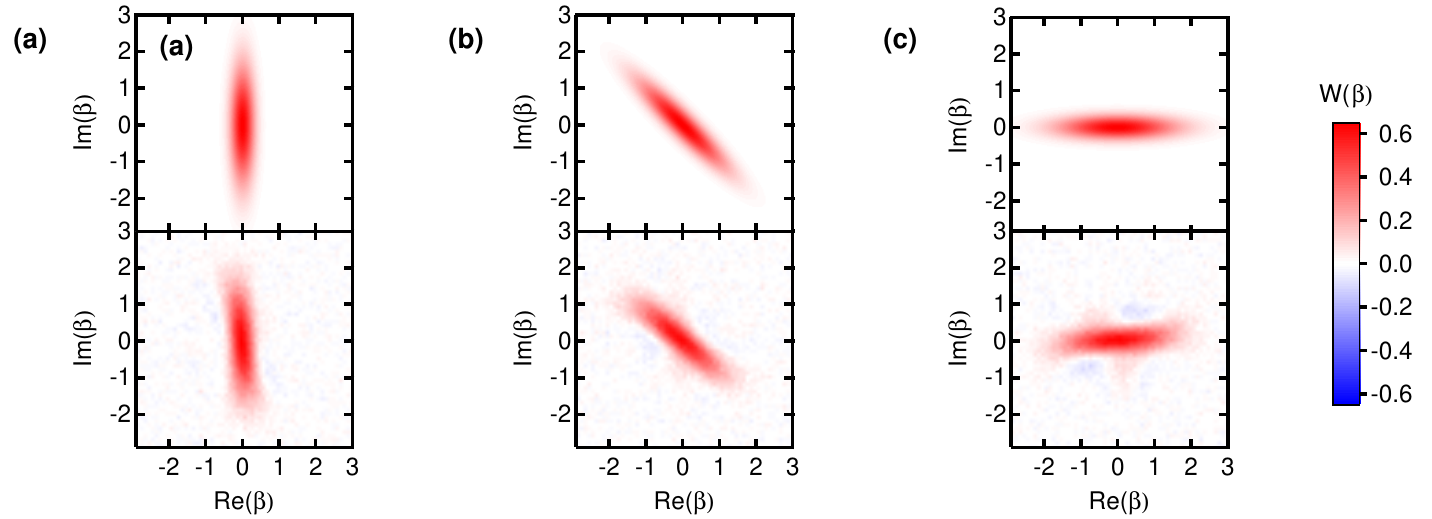}
\caption{(Color online) \textbf{Wigner tomography of the squeezed states.} Reconstruction of the squeezed states defined by Eq.~\ref{eq:squeezed_state} for $\xi =0.8$ (a), $0.8i$ (b), and $-0.8$ (c). We take the cutoff in the Fock-state expansion to be $N=8$. The calculated and measured Wigner functions are displayed in the upper and lower rows, respectively.}
\label{fig:squeezed_state} \vspace{-6pt}
\end{figure*}

In Fig.~\ref{fig:fock}, we also display the real and imaginary parts
of the density matrices in the Hilbert space obtained from the normalized
measured Wigner functions (the third and fourth rows, respectively).
As expected, the first $N+1$ photon number states are almost equally
populated, while the others remain unpopulated. The quantum coherence
of the superposed Fock states is manifested in the off-diagonal elements
$\rho_{m,n}$ with $m\neq n$. The non-vanishing imaginary parts of
the off-diagonal elements are mainly due to the imperfect setting
of the relative phases of different driving microwave components,
and they can be minimized further by more careful optimization. The
value of the matrix element $\rho_{m,n}$ ($0\leq m,n\leq N$) slightly
decreases as $m$ and $n$ increases since the decay rate is proportional
to $(m+n)/2$. The results further show that the produced states are
in good agreement with the ideal ones, demonstrating high-fidelity
generation of the desired states. 


\begin{table}[t]
\caption{Amplitudes $\alpha$ of the initial coherent states, Wigner function
reduction factors $R$, success probabilities $P$, and fidelities
$F$ of the produced truncated phase states with $\theta_{\mathrm{N,k}}=0$ (Fig.~\ref{fig:fock}).
The uncertainties are determined by bootstrapping on the measured
Wigner functions~\cite{VlastakisBell}. Bootstrapping allows random sampling with replacement at each pixel point of the measured Wigner function to obtain a pool of new Wigner functions that are used to extract $R$ and $F$ for a statistical analysis.}
\begin{tabular*}{0.48\textwidth}{@{\extracolsep{\fill}}@{\extracolsep{\fill}}cccc}
\hline
N, $k$ & 5, 0 & 6, 0 & 7, 0\tabularnewline
\hline
$|{\alpha}|$ & 1.63 & 1.74 & 1.85\tabularnewline
$R$ & $0.84\pm0.01$ & $0.82\pm0.01$ & $0.80\pm0.01$\tabularnewline
$P$ & $0.37$ & $0.31$ & $0.23$\tabularnewline
$F$ & $0.97\pm0.01$ & $0.96\pm0.01$ & $0.96\pm0.01$\tabularnewline
\hline
\end{tabular*} \vspace{-6pt}
\label{T:fock}
\end{table}

\begin{table}[t]
\caption{$R$, $P$, $F$, and $\langle (\Delta X(\theta/2))^2\rangle$ for the produced squeezed states with $r=0.8$ (Fig.~\ref{fig:squeezed_state}). All have an initial coherent state amplitude $|\alpha|=1.45$.}
\begin{tabular*}{0.48\textwidth}{@{\extracolsep{\fill}}@{\extracolsep{\fill}}cccc}
\hline
$\theta$ & 0 & $\pi$/2 & $\pi$ \tabularnewline
\hline
$R$ & $0.82\pm0.01$ & $0.84\pm0.01$ & $0.82\pm0.01$\tabularnewline
$P$ & $0.15$ & $0.14$ & $0.14$\tabularnewline
$F$ & $0.96\pm0.01$ & $0.94\pm0.01$ & $0.96\pm0.01$\tabularnewline
$\langle (\Delta X(\theta/2))^2\rangle$ & $0.067\pm0.001$ & $0.070\pm0.001$ & $0.070\pm0.001$ \tabularnewline
\hline
\end{tabular*} \vspace{-6pt}
\label{T:squeezed_state}
\end{table}



As another example to demonstrate the ability to generate any superposition state, we produce the squeezed state
\begin{equation}
\left| \xi \right\rangle =\frac{1}{\sqrt{\cosh r}}\sum_{n=0}^\infty \frac{(-e^{i\theta
}\tanh r)^n\sqrt{(2n)!}}{n!2^n}\left| 2n\right\rangle,
\label{eq:squeezed_state}
\end{equation}
with the squeeze parameter $\xi=re^{i\theta}$. For such a state the fluctuation of the quadrature $X(\theta /2)=(a^{\dagger }e^{i\theta /2}+ae^{-i\theta /2})/2$ is reduced by a factor $e^{-2r}$ compared to the vacuum level, so that it can be used for high-precision measurement~\cite{Caves1980}. The squeezed state can be well approximated by truncating the expansion into a superposition of the first $N+1$ Fock states, with $N$ depending on $r$. We approximately produce the squeezed states for $\xi =0.8$, $0.8i$, and $-0.8$ with a cutoff of $N=8$. The good agreement in the calculated and measured Wigner functions (Fig.~3) indicates the full controllability of both the amplitude and phase associated with each Fock state component. Table~\ref{T:squeezed_state} shows the corresponding $R$, $P$, $F$, and $\langle (\Delta X(\theta/2))^2\rangle$ [obtained from the Gaussian fit of the probability distribution of $X(\theta/2)$ calculated by integrating the measured Wigner function over $X(\theta/2+\pi/2)$] in the produced states. The measured $\langle (\Delta X (\theta/2)) ^2\rangle \ll 1/4$, showing that the initial quasiclassical states have been converted into quantum states with strong squeezing.

We note that, in the effective Hamiltonian of Eq.~\ref{eq:Hamiltonian}, we have ignored the cross-Kerr terms between the cavity and the readout cavity, $-\chi_{cr}a^{\dagger}aa_{r}^{\dagger}a_{r}$,
where $a_{r}^{\dagger}$ and $a_{r}$ are the creation and annihilation
operators of the readout cavity. This effect can either be corrected
by properly setting the initial phase of $\left|\alpha\right\rangle $
or the qubit driving phases. We have also ignored higher order terms,
among which the dominant terms are the cavity's self-Kerr effect $-Ka^{\dagger2}a^{2}/2$
and the correction to the dispersive qubit-cavity coupling $\chi_{qc}^{^{\prime}}a^{\dagger2}a^{2}\left|e\right\rangle \left\langle e\right|/2$.
Each of these terms causes additional photon-number-dependent phase
shifts; however, the resulting effect is small, as evidenced by the
high-fidelity of our experimental result.

In summary, we have proposed and experimentally demonstrated a new
single-step scheme to produce arbitrary superposition of Fock states
for a cavity. We show that a coherent state can be conditionally converted
to any desired superposition state by tailoring the probability amplitudes
associated with the superposed Fock state components. This is achieved
based on the experimentally convenient dispersive coupling between
the cavity and the qubit. The success probability of the scheme depends on the desired state. The achieved high fidelity of the experimental states prepared with this method opens up interesting applications of such non-classical states for implementation of quantum information and precision measurements. Of particular interest are the demonstrated truncated phase states that are directly applicable in quantum walk, which allows exponential speedup over classical computation for certain problems~\cite{Childs2003} and more importantly, can be used as a primitive for universal quantum computation~\cite{Childs2009}.

\begin{acknowledgments}
We acknowledge Zhangqi Yin, Andrei Petrenko, Brian Vlastakis, and Haidong Yuan for helpful discussions, and thank Tanay Roy and Madhavi Chand for the help with the parametric amplifier measurements. This work was supported by the Ministry of Science and the Ministry of Education of China through its grant to Tsinghua University, the National Natural Science Foundation of China under Grant Nos. 11674060 and 11474177, the Major State Basic Research Development Program of China under Grant No.2012CB921601, and the 1000 Youth Fellowship program in China.
\end{acknowledgments}


\end{document}


\title{Supplementary Material for ``Converting quasiclassical states into arbitrary Fock state superpositions in a superconducting cavity"}

\author{W.~Wang}
\author{L.~Hu}
\author{Y.~Xu}
\author{K.~Liu}
\author{Y.~Ma}
\affiliation{Center for Quantum Information, Institute for Interdisciplinary Information
Sciences, Tsinghua University, Beijing 100084, China}
\author{Shi-Biao Zheng}
\affiliation{Department of Physics, Fuzhou University, Fuzhou 350116, China}
\author{R.~Vijay}
\affiliation{Department of Condensed Matter Physics and Materials Science, Tata
Institute of Fundamental Research, Homi Bhabha Road, Mumbai 400005,
India}
\author{Y.~P.~Song}
\affiliation{Center for Quantum Information, Institute for Interdisciplinary Information
Sciences, Tsinghua University, Beijing 100084, China}
\author{L.-M. Duan}
\affiliation{Center for Quantum Information, Institute for Interdisciplinary Information
Sciences, Tsinghua University, Beijing 100084, China}
\affiliation{Department of Physics, University of Michigan, Ann Arbor, Michigan
48109, USA}
\author{L.~Sun}
\affiliation{Center for Quantum Information, Institute for Interdisciplinary Information
Sciences, Tsinghua University, Beijing 100084, China}
 \maketitle

\section{Device parameters}
\begin{figure}
\includegraphics{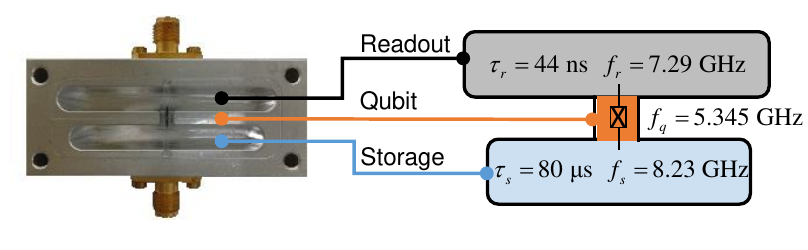}
\caption{\textbf{Experimental device.} A cross section and diagram of the device, which contains two 3D Al cavities and a sapphire chip with a transmon qubit located in a trench. The transmon qubit is strongly coupled to both cavities.}
\label{fig:device}
\end{figure}

Our experimental device consists of two waveguide cavities strongly coupled to a transmon qubit which is lithographically patterned on a sapphire chip, as shown in Fig.~\ref{fig:device}. One cavity called storage cavity is used for the preparation, manipulation, and storage of the photonic state, with a transition frequency $\omega_s/2\pi=8.230$~GHz and a decay rate $\kappa_s/2\pi=2$~kHz. The other cavity noted as the readout cavity with a transition frequency $\omega_r/2\pi=7.290$~GHz and a decay rate $\kappa_r/2\pi=3.62$~MHz, is used to detect the state of the qubit. The transmon qubit has a transition frequency $\omega _q/2\pi =5.345$ GHz with an anharmonicity $\alpha _q/2\pi=(\omega _{ge}-\omega _{ef})/2\pi =246$~MHz, an energy relaxation time $T_1=13.0~\mu$s and a pure dephasing time $T_\phi =13.8~\mu$s. The strong couplings between the qubit and the cavities result in state-dependent frequency shifts  $\chi _{qs}/2\pi =-1.44$~MHz (between qubit and storage cavity) and $\chi _{qr}/2\pi =-4.71$~MHz (between qubit and readout cavity). For simplicity, we refer to the storage cavity as ``the cavity'' henceforth.

\section{Readout properties}
In order to have a high-fidelity single-shot readout, besides the high-electron-mobility-transistor amplifier at 4K, we have a Josephson parametric amplifier (JPA)~\cite{Hatridge,Roy} connected to the output of the readout cavity as the first stage of amplification. The JPA works in a pulsed double-pumped mode~\cite{Kamal,Murch} to minimize pump leakage to the readout cavity. The schematic of the measurement setup can be found in Ref.~\onlinecite{Liu2016}. 

\begin{figure*}
\includegraphics{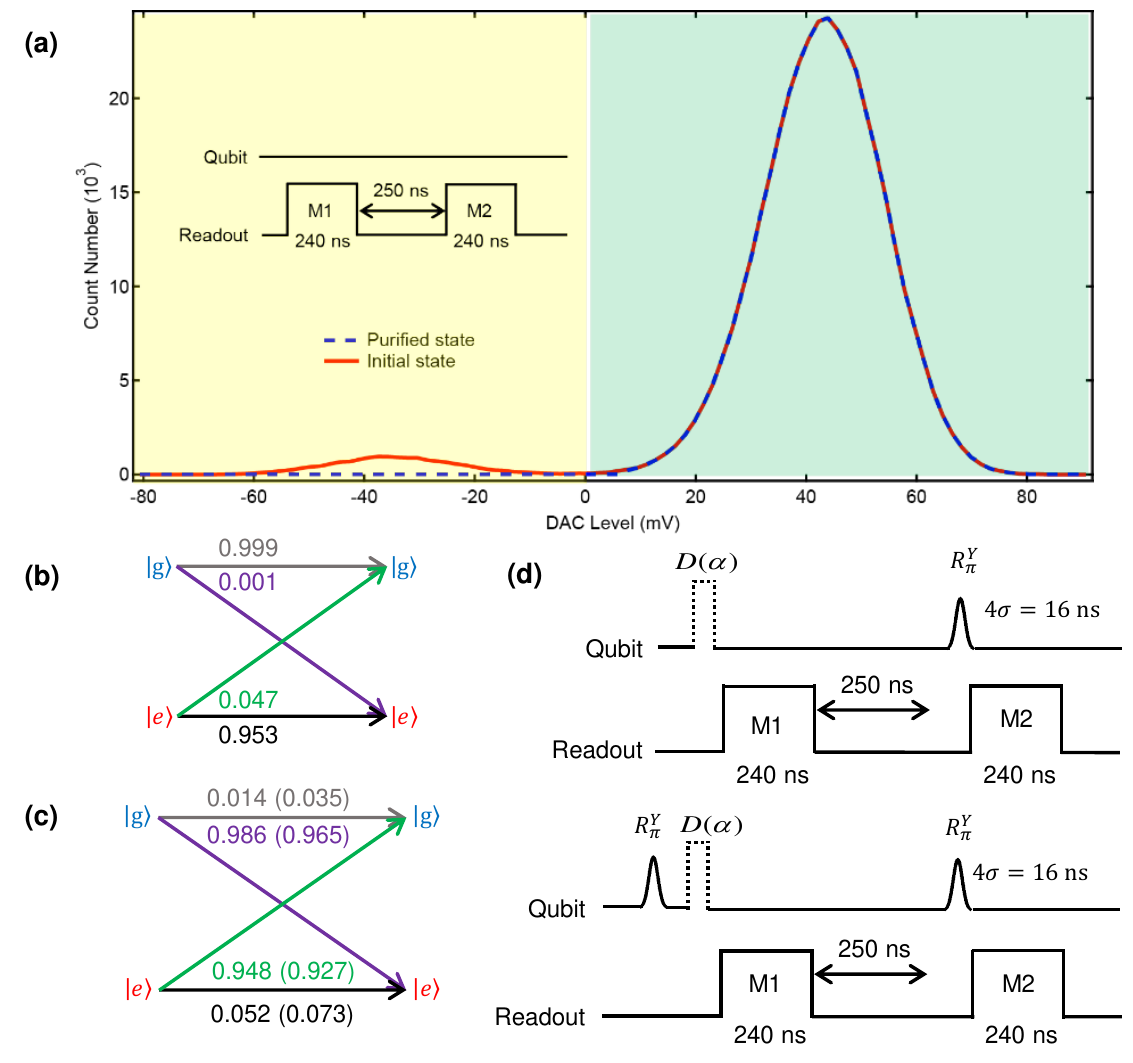}
\caption{\textbf{Readout properties.} \textbf{(a)} Bimodal and well-separated histogram of the qubit readout. A threshold $V_{th}=0$ is chosen to digitize the readout signal. The solid line corresponds to the case without the qubit purification measurement, showing the qubit has a 3.9\% probability of being populated in $\ket{e}$ state, while the dashed line represents the result after the qubit has been purified. The inset is the measurement pulse sequence. \textbf{(b)} Basic qubit readout matrix with the cavity left in vacuum. The readout errors are mainly due to the $T_1$ process. \textbf{(c)} Readout property of a $\pi$ pulse with a Gaussian envelope truncated to $4\sigma =16$~ns when the cavity is in the vacuum state. Numbers in parenthesis correspond to the case with $\bar{n}=3.5$ in the cavity. The loss of fidelity mainly due to qubit $T_1$ process during the measurement time and the waiting time between the two consecutive measurements. There is about an extra 2\% fidelity loss for the case with $\bar{n}=3.5$ photons in the cavity. \textbf{(d)} The pulse sequence used for the measurement in \textbf{(c)}.}
\label{fig:readout}
\end{figure*}

The characterizations of the readout properties of the qubit and cavity photon state are shown in Fig.~\ref{fig:readout}. Figure~\ref{fig:readout}a shows the histogram of the qubit readout. The histogram is clearly bimodal and well separated. A threshold $V_{th}=0$ is chosen to digitize the readout signal to $+1$ and $-1$ for the ground state $\ket{g}$ and the excited state $\ket{e}$, respectively. The qubit shows an unexpected excited state population of $3.9\%$ in the steady state, presumably due to stray infrared photons or other background noise leaking into the cavity. Together with the measured $T_1=13.0~\mu$s, we can derive $1/\Gamma_{\uparrow} \approx 330~\mu$s and $1/\Gamma_{\downarrow}\approx 13.5~\mu$s. We use measurement-based post-selection to purify the qubit to $\ket{g}$ state. After purification, the subsequent qubit measurement shows that the probability of the qubit being populated in $\ket{g}$ is as high as 0.999 (dashed histogram in Fig.~\ref{fig:readout}a), demonstrating the high quantum non-demolition property of the qubit readout. If we post-select the qubit at $\ket{e}$ state, the subsequent measurement finds the qubit remains in $\ket{e}$ with a fidelity of 0.953 (Fig.~\ref{fig:readout}b), with the errors dominantly from the $T_1$ process during the readout time (240~ns) and during the waiting time after the initialization measurement (250~ns). These readout properties are measured with the cavity left in vacuum.

Figure~\ref{fig:readout}c shows the property of a $\pi$ pulse ($\sigma=4$~ns) which is used to reset the qubit after generating the desired state by post-selection of the qubit at $\ket{e}$ state. Figure~\ref{fig:readout}d shows the associated measurement pulses. The numbers outside the parenthesis correspond to the cavity in the vacuum state, with the loss of fidelity mainly due to qubit $T_1$ process during the measurement time and the waiting time between the two consecutive measurements. The numbers inside the parenthesis correspond to the cavity in a coherent state with an average photon number of $\bar{n}=3.5$, with about 2\% lower fidelities due to the photon number occupations in the cavity.   

\begin{figure*}
\includegraphics{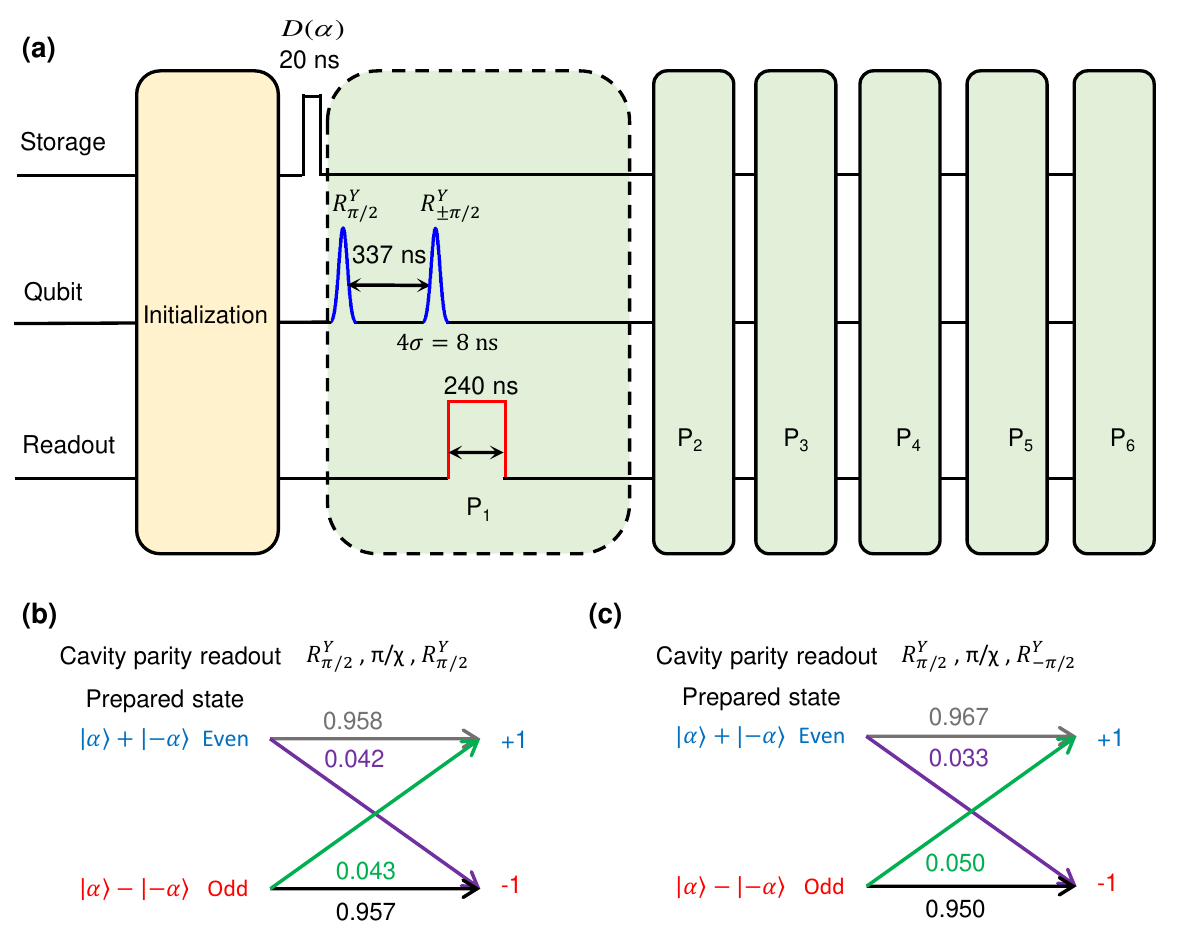}
\caption{\textbf{Parity readout properties.} \textbf{(a)} Protocol to measure parity readout fidelity. An initialization creates the $\ket{g,0}$ state, followed by a cavity displacement $D(\alpha=1)$ and six consecutive parity measurements. Post-selection of the first five consecutive identical parity results  would give a photon state parity with good confidence, and the sixth parity measurement is used for the final parity fidelity measurement. The pulse sequence for each parity measurement has been shown in $P_1$. \textbf{(b)} Parity readout property for given even and odd parity states for two protocols ( $R_{\pm\pi/2}^Y$ as the second qubit pulse) used in the tomography process. To minimize the system error in the tomography, we take both protocols and average the measurement results together.}
\label{fig:ParityFidelity}
\end{figure*}

The Wigner quasiprobability distribution $W(\beta)$ is measured by a combination of the cavity's displacement operation $D(-\beta)$ and a parity measurement, following a technique devised by Lutterbach and Davidovich~\cite{Lutterbach1997} and demonstrated in cavity QED~\cite{Bertet2002} and circuit QED~\cite{Vlastakis,SunNature,Liu2016}. The parity measurement is achieved in a qubit Ramsey-type measurement, where a conditional cavity $\pi$ phase shift $C(\pi)$ is sandwiched in between two unconditional qubit rotations $R_{\pi/2}^Y$. The parity readout property is shown in Fig.~\ref{fig:ParityFidelity}. The fidelity loss is dominantly from the qubit decoherence between the two $R_{\pi/2}^Y$ rotations. $T_1$ and $T_2$ processes together contribute about 3.8\% error.

\section{Measurement pulse sequence}
\begin{figure*}
\includegraphics{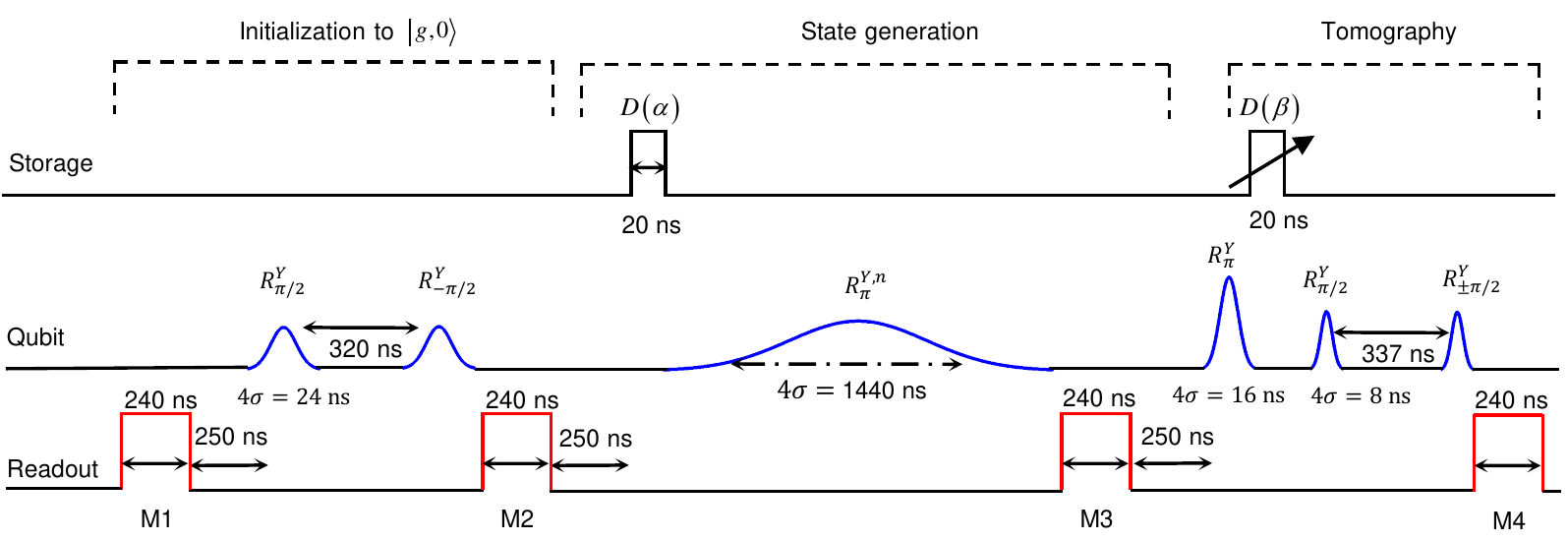}
\caption{\textbf{Schematic of the pulse sequence in the experiments.} The whole experiment can be divided into three main parts: 1. initialization of the system to $\ket{g,0}$ by post-selecting results of the qubit measurement and the cavity parity measurement; 2. creation of the desired arbitrary Fock-state superpositions by first displacing the cavity to $\ket{\alpha}$ state, then rotating the qubit simultaneously and independently in the subspaces \{$\ket{g,n}, \ket{e,n}$\}, and finally post-selecting the excited state of the qubit; 3. Wigner tomography measurement following a $\pi$ pulse reset of the qubit to the ground state.}
\label{fig:pulseSequence}
\end{figure*}

The pulse sequence for the experiment is shown in Fig.~\ref{fig:pulseSequence}. The whole experiment can be divided into three main parts: 1) initialization of the system to $\ket{g,0}$ by post-selecting results of the qubit measurement and the cavity parity measurement; 2) creation of the desired arbitrary Fock-state superpositions by first displacing the cavity to $\ket{\alpha}$ state, then rotating the qubit simultaneously and independently in the subspaces \{$\ket{g,n}, \ket{e,n}$\}, and finally post-selecting the excited state of the qubit; 3) Wigner tomography measurement following a $\pi$ pulse reset of the qubit to the ground state. The rotation of qubit $R_{e_n|n}(\beta _n)$ (already defined in the main text) in each subspace \{$\ket{g,n}, \ket{e,n}$\} is appropriately chosen depending on the probability amplitude of Fock state $\ket{n}$ in the desired superposition state.

The classical microwave pulse to displace the cavity state has a narrow square envelop with a width of 20~ns. The amplitude has been calibrated carefully as shown in the section of ``calibration of driving amplitude". All qubit drive pulses have a Gaussian envelope truncated to $\pm 2\sigma$. To eliminate the possible qubit leakage to higher qubit levels, we also apply the so-called ``derivative removal by adiabatic gate" technique~\cite{Motzoi} for pulses with $\sigma =2, 4$, and 6~ns.

\begin{figure*}[t]
\includegraphics{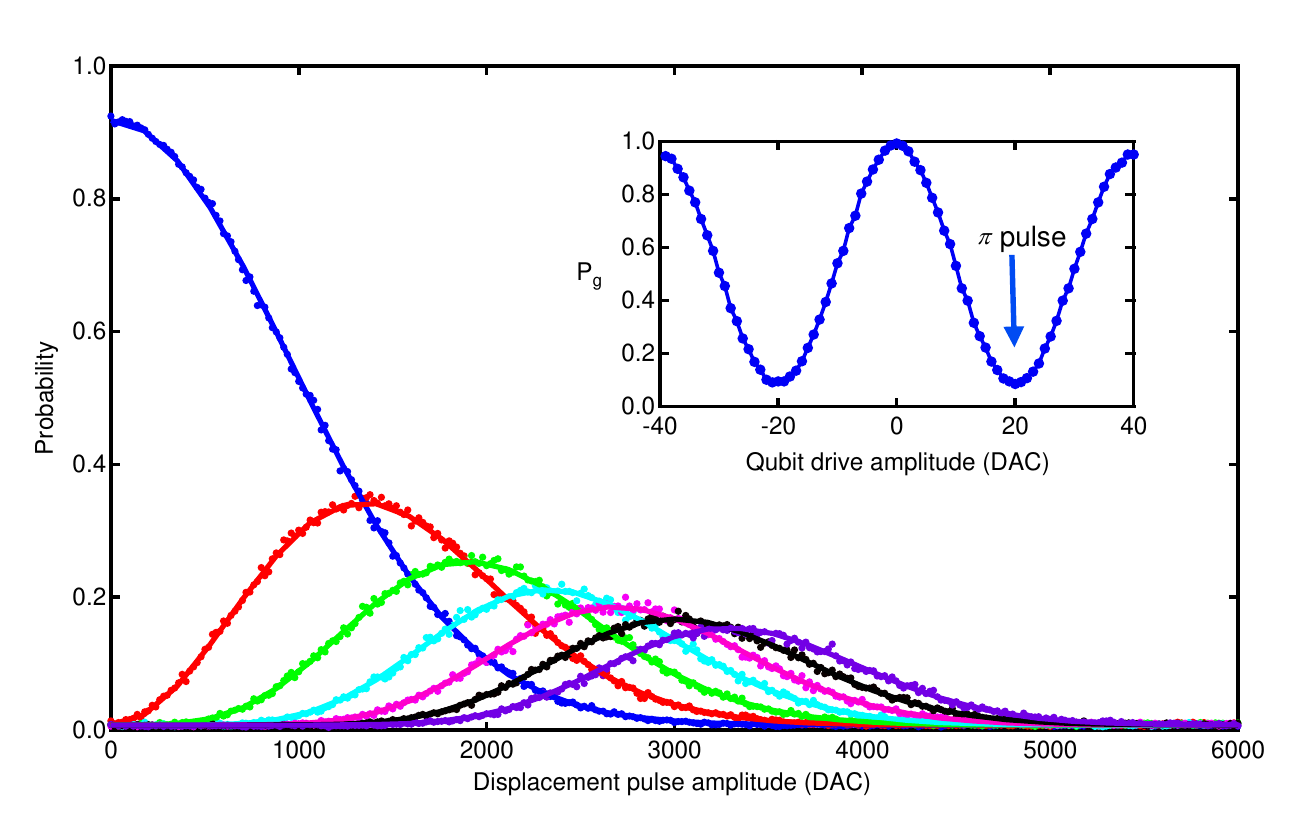}
\caption{\textbf{Poisson distribution of photon numbers in the cavity.} Dots are data for the probability of the first seven Fock states $N = 0,1,2,...6$ as a function of the displacement pulse amplitude but with a fixed width 20~ns. The probability measurement is performed with a selective $\pi$ pulse on the qubit at each resonant transition frequency corresponding to a specific photon number $n$ in the cavity. There is no normalization of the measurement data and the loss of probability dominantly comes from the $T_1$ process in the long duration of the $\pi$ pulse. Lines are from a global fit with a Poisson distribution $\displaystyle P(\ket{\alpha},N)=A|\alpha|^{2N}e^{-|\alpha|^2}/N!$, where $A$ is a scaling factor accounting for the probability loss. The excellent agreement indicates good control of the coherent state in the cavity. The displacement amplitude $\alpha$ is proportional to the measurement pulse amplitude. In this case, DAC level=1343 is calibrated to correspond to $\alpha=1$. Inset: Rabi oscillation to calibrate the qubit drive amplitude ($\sigma=360$~ns). The loss of contrast comes from the qubit relaxation in the long pulse duration.}
\label{fig:PoissonDist}
\end{figure*}

\section{Calibration of driving amplitude}
The calibration of the qubit driving amplitude is essential in creating a high-fidelity Fock-state superposition with our protocol. This calibration is performed by a typical Rabi oscillation with a selective driving pulse ($\sigma=360$~ns) at each resonant transition frequency of the qubit corresponding to a specific photon number $n$ in the cavity (inset of Fig.~\ref{fig:PoissonDist}). The qubit decoherence during the long rotation only reduces the Rabi signal contrast without affecting the calibration of the driving amplitude with a deviation typically less than 1\%.

The calibration of the driving amplitude to create a desired coherent state in the cavity is also crucial. This calibration is performed by measuring the Poisson distribution of photon numbers in the cavity (Fig.~\ref{fig:PoissonDist}). We measure the probability of the first seven Fock states $N = 0,1,2,...6$ through a selective $\pi$ pulse ($\sigma=360$~ns) as a function of the displacement pulse amplitude but with a fixed width of 20~ns. A nearly perfect global fit to the measurement results with a Poisson distribution not only gives the required calibration of the cavity displacement amplitude, but also indicates good control of the coherent state in the cavity.

\begin{table*} [t]
\caption{Error budgets for the final Wigner tomography used to examine the produced states. These error sources in total account for the large reduction of $R$.} 
\begin{tabular*}{0.96\textwidth}{@{\extracolsep{\fill}}@{\extracolsep{\fill}}cccc}
\hline
  source of errors & duration ($\mu$s) & coherence time ($\mu$s) & contribution to reduction of $R$ from 1 \\\hline
  qubit $T_1$ between M3 and $\pi$ pulse & $\sim0.370$ & 13.0 & 5.7$\%$ \\
  qubit $T_1$ between two $\pi/2$ pulses in tomography& 0.337 & 13.0 & 2.6$\%$ \\
  qubit $T_\phi$ between two $\pi/2$ pulses in tomography & 0.337 & 13.8 & 4.9$\%$ \\
  inaccuracy of $\pi$ pulse & 0.016 & NA & $2\% -4\%$ \\
  cavity decay between two $\pi/2$ pulses in tomography & 0.337 & $80/\bar{n}$ &  $\sim1\%$ \\
\hline
\end{tabular*} \vspace{-6pt}
\label{T:ErrorBudgets}
\end{table*}

\section{Analysis of errors}
Our selective pulse is relatively slow (about 1.44~$\mu$s) due to the small dispersive frequency shift, however, to generate the truncated phase states and the squeezed states presented in our main text our scheme is not sensitive to the qubit energy relaxation $T_1$, qubit dephasing $T_\phi$, and cavity photon energy relaxation. First of all, we apply microwave drives to qubit only associated with Fock states present in the target state and only select the qubit's excited state. The probability of selecting unwanted Fock states in the final state (orthogonal to the desired state) can be neglected because the probability of  $\Gamma_{\uparrow}$ process ($\ket{g}$ to $\ket{e}$) is only about 1.44$\mu$s/330$\mu$s =0.4\%.
 

Secondly, random $T_1$ decay from $\ket{e}$ to $\ket{g}$ during the photon-number-dependent qubit rotation obviously decreases the probability of the qubit being excited; however, the fidelity of the produced state is insensitive to the qubit energy decay. This has a simple explanation: The purity of the produced state is insensitive to the qubit decay as this process cannot mix the cavity state correlated with the qubit's ground state into that with the qubit's excited state; instead, this decay mainly leads to slight changes of the coefficients of the superposed Fock states, to which the fidelity of the produced state is robust. A numerical simulation shows this $T_1$ decay reduces the fidelity by only 1-3\%. On the other hand, the qubit dephasing corresponds to a random qubit frequency fluctuation, which just results in slight changes of the probability amplitudes associated with the Fock state components of the produced state. The fidelity of the produced state is insensitive to such slight changes and thus robust against the qubit dephasing. This is confirmed by numerical simulation, which shows that the fidelity $F$ of the produced cavity state is even more insensitive to the qubit dephasing than to the qubit energy relaxation.

Thirdly, we consider the effect from the cavity photon decay. This effect increases with the photon number (as shown in Fig.~2 of the main text, the matrix element $\rho_{m,n}$ has a smaller value for larger $m$ and $n$), however, numerical simulations show that with our long photon lifetime ($80~\mu$s), photon decay actually contributes less to the fidelity loss than the qubit decoherence for the generated states in the main text.

Therefore, after considering all the above contributions to the fidelity loss, the overall fidelity could still be as high as 96\%.

In contrast, both $T_1$ and $T_\phi$ process cause a much more significant reduction of $R$ in the following Wigner tomography to reconstruct the cavity state. The error budgets are shown in Table~\ref{T:ErrorBudgets}. Explicitly, 1) the qubit $T_1$ process between the final detection of $\ket{e}$ to generate the target state and the tomography causes a reduction of 0.370$\mu$s/13$\mu$s$\times$2=5.7\%. Here 0.370~$\mu$s includes half of the measurement time of $M_3$ to estimate the decoherence during the measurement, and the factor two comes from the fact that parity is from -1 to 1. 2) Qubit $T_1$ process during the parity measurement causes a reduction of 0.337$\mu$s/13$\mu$s=2.6\%. 3) Qubit dephasing $T_\phi$ during the parity measurement causes a reduction of 0.337$\mu$s/13.8$\mu$s$\times$2=4.9\%. 4) Due to the inaccuracy of the $\pi$ pulse (calibrated when the cavity is in the vacuum state), the qubit has a probability of 1-2\% to remain in the exited state $\ket{e}$ for large $N$ due to the dispersive frequency shift (see Fig.~\ref{fig:readout}c), leading to a reduction of 2-4\%. 5) We note that the cavity photon decay only contributes to the reduction of $R$ during the dispersive qubit-cavity interaction sandwiched in between the final two Ramsey pulses. This contribution is roughly estimated as $\bar{n}\times$0.337$\mu$s/80$\mu$s$\sim$1\%. With all of the effects being included, the resulting Wigner function reduction factor is about 0.83, in agreement with the measured results.